\documentclass{ws-procs9x6}

\begin{document}

\title{Higher Twist Effects in Polarized DIS
\footnote{\uppercase{T}his research was supported by the
\uppercase{UK} \uppercase{R}oyal \uppercase{S}ociety and
\uppercase{JINR}-\uppercase{B}ulgaria \uppercase{C}ollaborative
\uppercase{G}rants and by the \uppercase{RFBR} (No 05-01-00992,
03-02-16816).}}

\author{E. Leader}

\address{Imperial College London\\ Prince Consort Road, London SW7 2BW,
England\\
E-mail: e.leader@imperial.ac.uk}

\author{ A.~V. Sidorov}

\address{Bogoliubov Theoretical Laboratory\\
Joint Institute for Nuclear Research, 141980 Dubna, Russia\\
E-mail:sidorov@thsun1.jinr.ru }

\author{D.~B. Stamenov}

\address{Institute for Nuclear Research and Nuclear Energy\\
Bulgarian Academy of Sciences\\
Blvd. Tsarigradsko Chaussee 72, Sofia 1784, Bulgaria \\
E-mail: stamenov@inrne.bas.bg}

\maketitle

\abstracts{The size of higher twist corrections to the spin proton
and neutron $g_1$ structure functions and their role in
determining the polarized parton densities in the nucleon is
discussed. }

\section{Introduction}

The study of the quark-hadron duality phenomena\cite{BG} using
the present much more precise data for the polarized and
unpolarized structure functions is in a progress\cite{BFL}. The
better understanding of the non-perturbative higher twist effects
is important for this analysis, especially in the polarized case,
where the investigations are in the very beginning.

In this talk we will discuss both, the size of the higher twist
effects in {\it polarized} DIS as well as their role in the
determination of the polarized parton densities (PPDs) in the
nucleon using different approaches of QCD fits to the data.

\section{QCD Treatment of the $g_1$ Structure Function}

In QCD the spin structure function $g_1$ can be written in the
following form ($Q^2 >> \Lambda^2$):
\begin{equation}
g_1(x, Q^2) = g_1(x, Q^2)_{\rm LT} + g_1(x, Q^2)_{\rm HT}~,
\label{g1QCD}
\end{equation}
where "LT" denotes the leading twist ($\tau=2$) contribution to
$g_1$, while "HT" denotes the contribution to $g_1$ arising from
QCD operators of higher twist, namely $\tau \geq 3$. In Eq.
(\ref{g1QCD}) (the nucleon target label N is dropped)
\begin{equation}
g_1(x, Q^2)_{\rm LT}= g_1(x, Q^2)_{\rm pQCD} + h^{\rm TMC}(x,
Q^2)/Q^2 + \mathcal {O}(M^4/Q^4)~,
\label{g1LT}
\end{equation}
where $h^{\rm TMC}(x, Q^2)$ are the calculable{\cite{TMC}
kinematic target mass corrections, which effectively belong to the
LT term. $g_1(x, Q^2)_{\rm pQCD}$ is the well known (logarithmic
in $Q^2$) pQCD expression and in NLO has the form
\begin{equation}
g_1(x,Q^2)_{\rm pQCD}={1\over 2}\sum _{q} ^{N_f}e_{q}^2 [(\Delta q
+\Delta\bar{q})\otimes (1 + {\alpha_s(Q^2)\over 2\pi}\delta C_q)
+{\alpha_s(Q^2)\over 2\pi}\Delta G\otimes {\delta C_G\over N_f}],
\label{g1partons}
\end{equation}
where $\Delta q(x,Q^2), \Delta\bar{q}(x,Q^2)$ and $\Delta
G(x,Q^2)$ are quark, anti-quark and gluon polarized densities in
the proton, which evolve in $Q^2$ according to the spin-dependent
NLO DGLAP equations. $\delta C(x)_{q,G}$ are the NLO
spin-dependent Wilson coefficient functions and the symbol
$\otimes$ denotes the usual convolution in Bjorken $x$ space. $\rm
N_f$ is the number of active flavors.

In Eq. (\ref{g1QCD})
\begin{equation}
g_1(x, Q^2)_{\rm HT}= h(x, Q^2)/Q^2 + \mathcal {O}(1/Q^4)~,
\label{HTQCD}
\end{equation}
where $h(x, Q^2)$ are the {\it dynamical} higher twist ($\tau=3$
and $\tau=4$) corrections to $g_1$, which are related to
multi-parton correlations in the nucleon. The latter are
non-perturbative effects and cannot be calculated without using
models. That is why a {\it model independent} extraction of the
dynamical higher twists $h(x, Q^2)$ from the experimental data is
important not only for a better determination of the polarized
parton densities but also because it would lead to interesting
tests of the non-perturbative QCD regime and, in particular, of
the quark-hadron duality.

One of the features of polarized DIS is that a lot of the present
data are in the preasymptotic region ($Q^2 \sim 1-5~GeV^2,
4~GeV^2 < W^2 < 10~GeV^2$). While in the unpolarized case we can
cut the low $Q^2$ and $W^2$ data in order to minimize the less
known higher twist effects, it is impossible to perform such a
procedure for the present data on the spin-dependent structure
functions without losing too much information. This is especially
the case for the HERMES, SLAC and Jefferson Lab experiments. So,
to confront correctly the QCD predictions with the experimental
data and to determine the {\it polarized} parton densities
special attention must be paid to the non-perturbative higher
twist (powers in $1/Q^2$) corrections to the nucleon structure
functions.

\section{Higher Twists and Their Role in Determining PPDs}

We have used two approaches to extract the polarized parton
densities from the world polarized DIS data. According to the
first\cite{LSS2001} the leading twist LO/NLO QCD expressions for
the structure functions $g_1$ and $F_1$ have been used in order
to confront the data on spin asymmetry $A_1 (\approx g_1/F_1)$
and $g_1/F_1$. We have shown\cite{LomConf,newHTA1} that in this
case the extracted from the world data 'effective' HT corrections
$h^{g_1/F_1}(x)$ to the ratio $g_1/F_1$
\begin{equation}
\left[{g_1(x,Q^2)\over
F_1(x,Q^2)}\right]_{exp}~\Leftrightarrow~{g_1(x,Q^2)_{\rm LT}\over
F_1(x,Q^2)_{\rm LT}} + {h^{g_1/F_1}(x)\over Q^2} \label{A1HT}
\end{equation}
are negligible and consistent with zero within the errors, {\it
i.e.} $h^{g_1/F_1}(x) \approx 0$, when for $(g_1)_{LT}$ and
$(F_1)_{LT}$ their NLO QCD approximations are used. (Note that in
QCD the unpolarized structure function $F_1$ takes the same form
as $g_1$ in (\ref{g1QCD}), namely $F_1 = (F_1)_{LT} +
(F_1)_{HT}.$) What follows from this result is that the higher
twist corrections to $g_1$ and $F_1$ approximately {\it
compensate} each other in the ratio $g_1/F_1$ and the NLO PPDs
extracted this way are less sensitive to higher twist effects.
This is not true in the LO case (see our discussion in Ref. 7).
The set of polarized parton densities extracted this way is
referred to as PD($g_1^{\rm NLO}/F_1^{\rm NLO}$).

According to the second approach\cite{LSSHT}, the $g_1/F_1$ and
$A_1$ data have been fitted using phenomenological
parametrizations of the experimental data for the unpolarized
structure function $F_2(x,Q^2)$ and the ratio $R(x,Q^2)$ of the
longitudinal to transverse $\gamma N$ cross-sections (i.e. $F_1$
is replaced by its expression in terms of usually extracted from
unpolarized DIS experiments $F_2$ and $R$). Note that such a
procedure is equivalent to a fit to $(g_1)_{exp}$, but it is more
consistent than the fit to the $g_1$ data themselves actually
presented by the experimental groups because here the $g_1$ data
are extracted in the same way for all of the data sets. In this
case the HT corrections to $g_1$ cannot be compensated because the
HT corrections to $F_1(F_2$ and $R)$ are absorbed in the
phenomenological parametrizations of the data on $F_2$ and $R$.
Therefore, to extract correctly the polarized parton densities
from the $g_1$ data, the HT corrections (\ref{HTQCD}) to $g_1$
have to be taken into account. So, according to this approach we
have used the following expression for the ratio $g_1/F_1$:
\begin{equation}
\left[{g_1^N(x,Q^2)\over F_1^N(x,
Q^2)}\right]_{exp}~\Leftrightarrow~ {{g_1^N(x,Q^2)_{\rm
LT}+h^N(x)/Q^2}\over F_1^N(x,Q^2)_{exp}}~,
\label{g1F2Rht}
\end{equation}
where $g_1^N(x,Q^2)_{\rm LT}$ (N=p, n, d) is given by the leading
twist expression (\ref{g1partons}) in LO/NLO approximation
including the target mass corrections. In (\ref{g1F2Rht}) $h^N(x)$
are the dynamical $\tau=3$ and $\tau=4$ HT corrections which  are
extracted in a {\it model independent way}. In our analysis their
$Q^2$ dependence is neglected. It is small and the accuracy of the
present data does not allow to determine it. The set of PPDs
extracted according to this approach is referred to as
PD($g_1^{\rm LT}+\rm HT$). The details of our recent analysis
using the present available data on polarized DIS are given in
\cite{LSS05}.

The extracted higher twist corrections to the proton and neutron
spin structure functions, $h^p(x)$ and $h^n(x)$, are shown in Fig.
1. As seen from Fig. 1 the size of
\begin{figure}[ht]
\centerline{\epsfxsize=2.2in\epsfbox{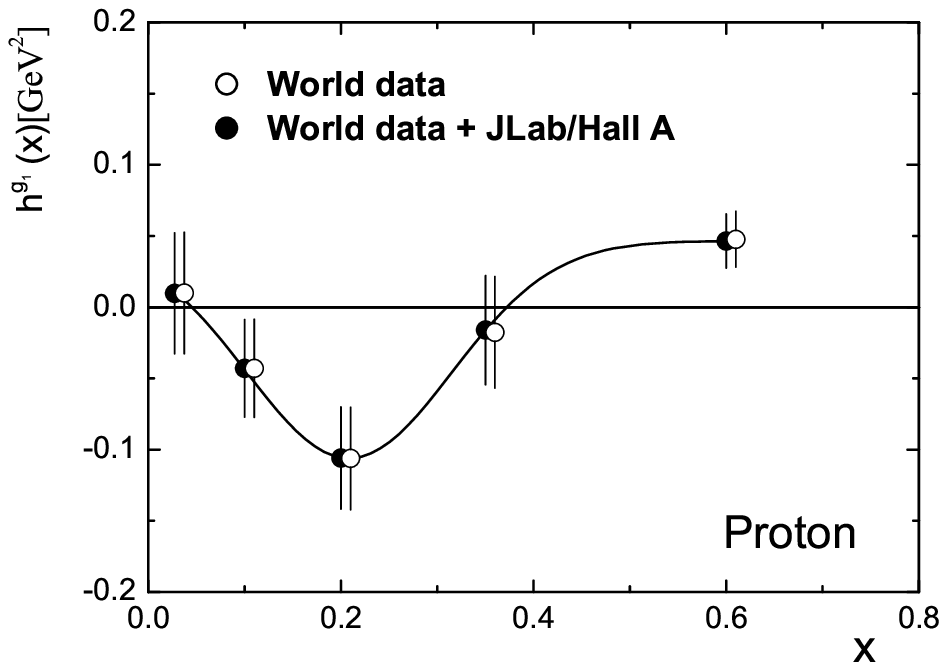}
\epsfxsize=2.2in\epsfbox{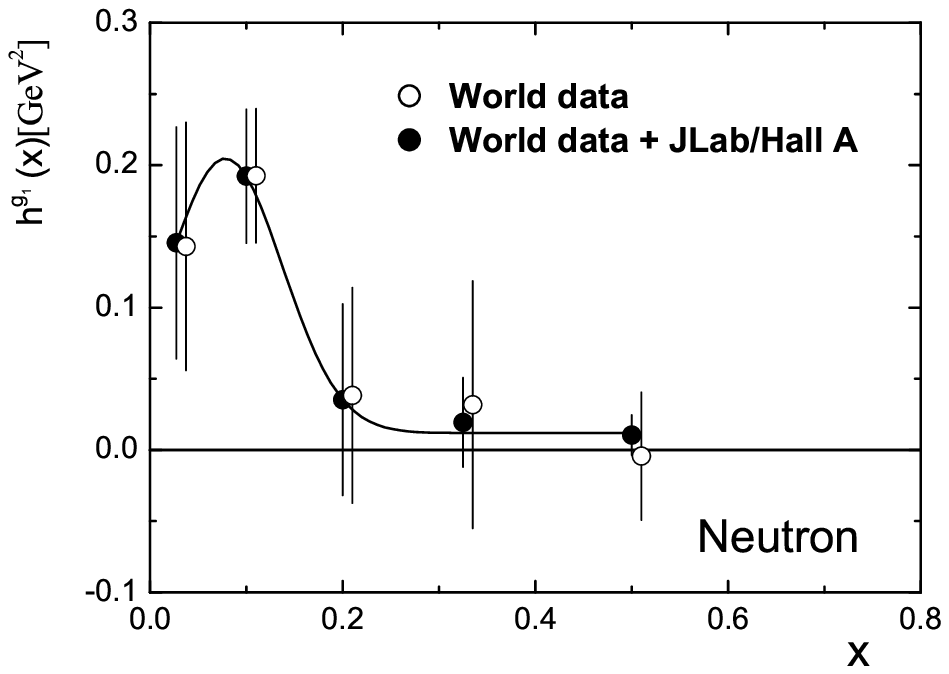} }
 \caption{
Higher twist corrections to the proton and neutron $g_1$ structure
functions extracted from the data on $g_1$ in NLO(${\rm \overline
{MS}}$) QCD approximation for $g_1(x,Q^2)_{\rm LT}$. The
parametrization (\ref{HTparam}) of the higher twist values is
also shown.\label{inter1}}
\end{figure}
the HT corrections is not negligible and their shape depends on
the target. In Fig. 1 our previous results on the higher twist
corrections to $g_1$ (before the JLab Hall A data were available)
are also presented. As seen from Fig. 1, thanks to the very
precise JLab Hall A data\cite{JLab} at large $x$ the higher twist
corrections to the neutron spin structure function are now much
better determined in this region. In Fig. 1 our parametrizations
of the values of higher twists for the proton and neutron targets

\begin{eqnarray}
\nonumber h^p(x)&=&0.0465 - {0.1913 \over \sqrt{\pi/2}}
exp[-2((x-0.2087)/0.2122)^2] \\
[2mm] h^n(x)&=&0.0119 + {0.2420 \over \sqrt{\pi/2}}
exp[-2((x-0.0783)/0.1186)^2] \label {HTparam}
\end{eqnarray}
are also shown. These should be helpful in a calculation of the
nucleon structure function $g_1$ for any $x$ and moderate $Q^2$ in
the experimental region, where the higher twist corrections are
not negligible. The impact of the very recent COMPASS
data\cite{COMPASS} on the values of higher twist corrections is
negligible. The new values are in a good agreement with the old
ones\cite{LSS05}.

The values of the higher twist corrections to the proton and
neutron $g_1$ structure functions extracted in a model independent
way from polarized DIS data are in agreement with the QCD sum rule
estimates\cite{Balitsky:1990jb} as well as with the instanton
model predictions\cite{Balla:1997hf} but disagree with the
renormalon calculations\cite{HTrenorm}. About the size of the HT
corrections extracted from the resonance region see the discussion
in the Fantoni's talk at this Workshop.

\begin{figure}[ht]
\centerline{ \epsfxsize=2.2in\epsfbox{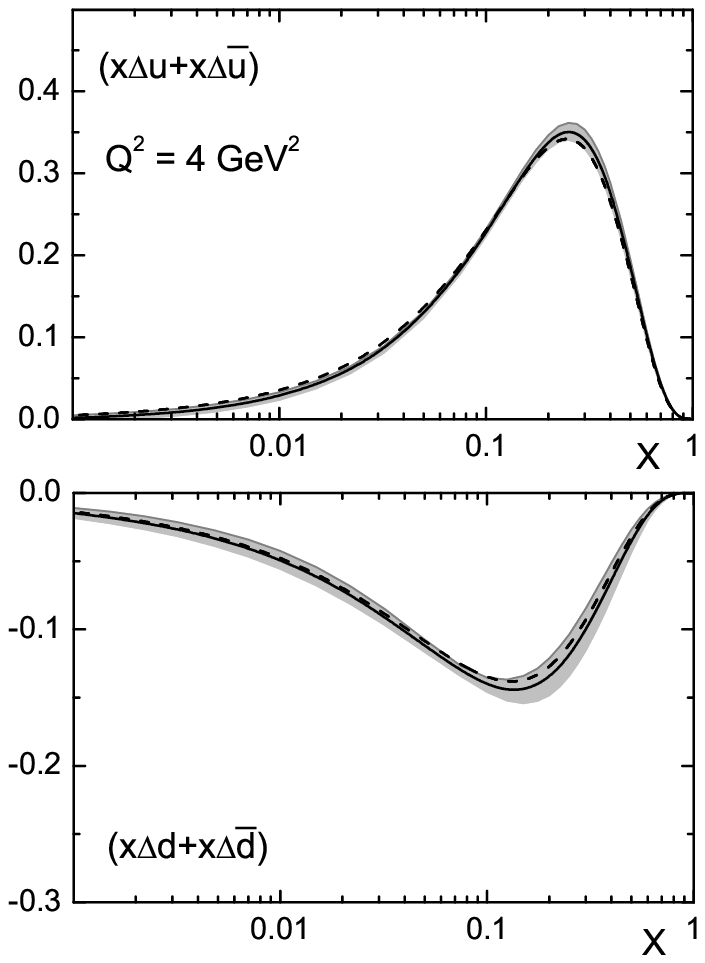}
\epsfxsize=2.2in\epsfbox{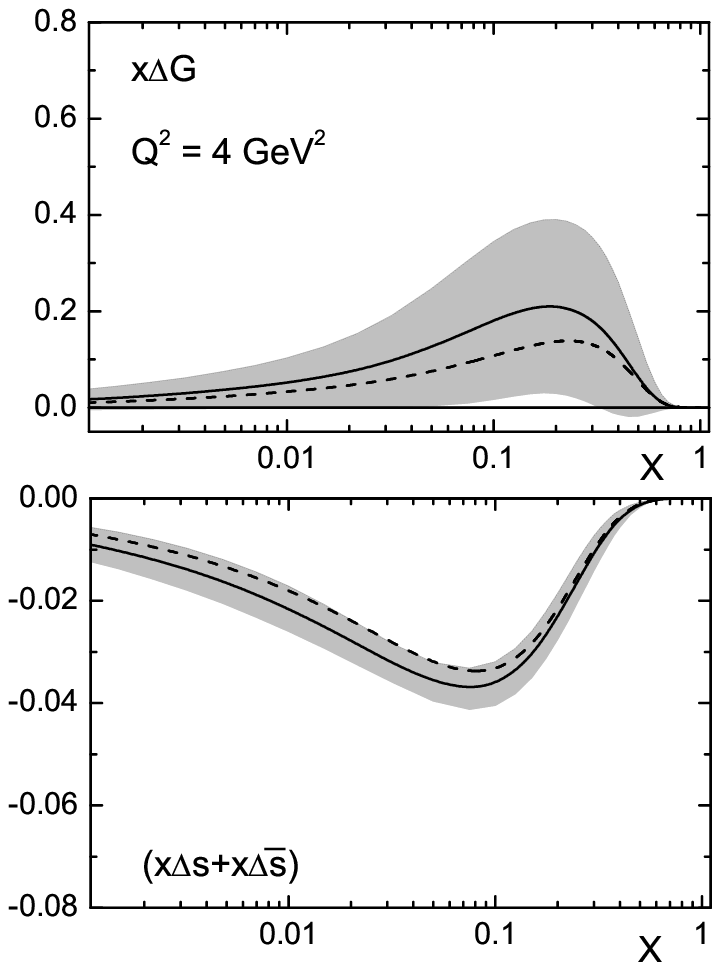} }
 \caption{NLO($\rm \overline{MS}$) polarized parton densities
PD($g_1^{\rm LT}+\rm HT$) (solid curves) together with their error
bands compared to PD($g_1^{\rm NLO}/F_1^{\rm NLO}$) (dashed
curves) at $Q^2=4~GeV^2$. \label{inter2}}
\end{figure}

In Fig. 2 we compare the NLO($\rm \overline{MS}$) polarized parton
densities PD($g_1^{\rm LT}+\rm HT$) with PD($g_1^{\rm
NLO}/F_1^{\rm NLO}$). As seen from Fig. 2 the two sets of PPDs are
very close to each other, especially for $u$ and $d$ quarks. This
is a good illustration of the fact that a fit to the $g_1$ data
taking into account the higher twist corrections to $g_1$
($\chi^2_{\rm DF,NLO}=0.872$) is equivalent to a fit of the data
on $A_1(\sim g_1/F_1~)$ and $g_1/F_1$ using for the $g_1$ and
$F_1$ structure functions their NLO leading twist expressions
($\chi^2_{\rm DF,NLO}=0.874$). In other words, this fact confirms
once more that the HT corrections to $g_1$ and $F_1$ approximately
cancel in the ratio $g_1/F_1$. Nevertheless, we consider that the
set of the polarized parton densities PD($g_1^{\rm LT}+\rm HT$) is
preferable because using them and simultaneously extracted higher
twist corrections to $g_1$, the spin structure function $g_1$ can
be correctly calculated in the preasymptotic $(Q^2,~W^2)$ region
too.\\

{\it In conclusion,} the higher twist effects in polarized DIS
have been studied. It was shown that the size of the HT
corrections to the spin structure function $g_1$ is not negligible
and their shape depends on the target. It was also demonstrated
that their role is important for the correct determination of the
polarized parton densities.

\end{document}